# Equilibrium SAT based PQC: New aegis against quantum computing

Keum-Bae Cho

**Affiliations:**

B.S. in Electrical Engineering at KAIST, South Korea; M.S. and Ph.D. in Electrical Engineering and Computer Science at Seoul National University, South Korea

*Corresponding author, E-mail (kbcho98@snu.ac.kr)

**Abstracts:**
Public-key cryptography algorithms have evolved towards increasing computational complexity to hide desired messages, which is accelerating with the development of the Internet and quantum computing. This paper introduces a novel public-key cryptography algorithm that generates ciphertexts by counting the number of elements in randomly extracted subsets from a multiset. After explaining the novel cryptographic concept, the process of mathematically refining it using satisfiability problems is described. The advantages of the proposed algorithm are: first, it is significantly faster than other public-key algorithms; second, it does not require big numbers, making it executable on any devices; and third, it can be easily extended into a public-key cryptosystem using a single public key and multiple private keys while maintaining quantum resistance.

**One Sentence Summary:** a new SAT based post-quantum cryptography that does not require complex computations is proposed.

## INTRODUCTION

The fundamental problems underlying public-key cryptographic algorithms—such as integer factorization and the discrete logarithm problem—can now be solved in polynomial time using quantum computers via Shor's algorithm *(1, 2)*. As quantum computer development technology continues to advance *(3-5)*, the security of public-key cryptographic algorithms based on these problems can no longer be guaranteed. Therefore, new public-key algorithms that do not rely on the difficulty of integer factorization or the discrete logarithm problem are required. Such cryptographic techniques are called quantum-resistant cryptography or Post-Quantum Cryptography (PQC). To date, quantum-resistant cryptographic algorithms, excluding hash-based algorithms used for digital signatures, are being researched in four main areas: multivariable-based, code-based, isogeny-based, and lattice-based.

The National Institute of Standards and Technology (NIST) have been conducting an algorithm selection process since 2017 to standardize post-quantum cryptographic algorithms. For the third round of standardization in the Public Key Encryption field, three lattice-based algorithms—CRYSTALS-Kyber*(6)*, NTRU*(7)*, and SABER*(8)*—and one code-based algorithm, Classic McEliece*(9)*, were selected as candidates. HQC*(10)* was added as a candidate for the fourth round, and ultimately CRYSTALS-Kyber and HQC were selected. The selected lattice-based and code-based algorithms add a random vector to impart a one-way property that makes it difficult to derive the private key from the public key. They also impose constraints on the random values added to ensure a trapdoor is created. These constraints provide the basis for creating various attack algorithms, leading to persistent attempts to obtain or recover private keys or original data.

This paper introduces a novel cryptographic concept that uses an array as ciphertext, created by randomly extracting subsets of a multiset and counting the number of elements contained in each. It then describes the process of mathematically formalizing this concept using the satisfiability problem (SAT). The proposed algorithm differs from existing approaches that add or multiply random values to create one-way properties; it instead employs a random extraction method.

Public-key cryptographic algorithms are based on NP-hard problems or problems predicted to be NP-hard *(11)*. SAT is not only the first proven NP-complete (NP-hard & $\in$ NP) problem *(12)*, but if the clauses constituting SAT can be used as the public key, it becomes immune to various algebraic attacks attempting to derive the private key from the public key. The security strength of the private key then depends solely on the performance of search algorithms like SAT solvers. An example proposed a cryptographic algorithm that transforms the components constituting the public key into Algebraic Normal Form (ANF), and then multiplies them by random ANFs corresponding to noise *(13)*. However, it was not proven that decoding the ciphertext within polynomial time to find the plaintext is impossible, and the encryption time and ciphertext size were too large for practical use. The algorithm proposed in this paper is based on the difficulty of finding a solution used in the process of generating a public key for enhanced SAT problems with strengthened solution conditions. In conventional SAT problems, a solution required at least one literal to be TRUE in every clause. However, in this paper, the conditions for a solution to k-SAT are strengthened: when k is even, a solution must have exactly k/2 literals that are TRUE in every clause; when k is odd, a solution must have exactly ⌊k/2⌋ or exactly ⌊k/2⌋+1 literals that are TRUE in every clause. We define the satisfiability problem with strengthened conditions as **equilibrium SAT**, meaning that the literals with TRUE values and those with FALSE values within a single clause are balanced.

# RESULTS

### Concept

Let the number of input variables be *n*, the multiplicity of the multiset be *m*, the number of elements in a subset be *2k*, and the number of randomly extracted subsets be *e*.

Prepare *mn* white balls and *mn* gray balls, and write numbers from *1* to *n* on each ball, with *m* balls bearing each number.

For each number, randomly select either white or gray. After placing gold on the selected color, record which color received gold on a notepad.

Separate the balls with gold and those without gold into two groups.

Then, randomly extract balls from both groups to form bundles of *2k* balls that satisfy the following two conditions:

First, bundle all balls so that each bundle contains exactly *k* gold balls.

Second, ensure each bundle contains *2k* balls with distinct numbers.

Place the generated bundles into a basket. Randomly select *e* bundles. Starting from the smallest number, count the number of balls with the same number written on them. Record this as <white count, gray count> to create an array like the following:

[<4,5>, **<5,3>**, …, <5,5>]

Additionally, randomly select pairs where the white count and gray count differ, creating another array where the white and gray counts are swapped, as shown below:

[<4,5>, **<3,5>**, …, <5,5>]

The array with swapped counts is defined as the contaminated array.

Colors were randomly selected to place gold inside the balls. The number of white balls and gray balls is identical. The criterion for forming groups is not color, but the number of balls containing gold. Therefore, the distribution of colors across the bundles exhibits randomness. Furthermore, since the bundles were randomly extracted and then opened to count the number of balls bearing the same number, the randomness characteristic propagates to the individual numbers, affecting both the number of white balls and the number of gray balls. Consequently, to distinguish between a normal array and a contaminated array, one must analyze all pairs of numbers composing the array, making it difficult to easily discern the difference between the two arrays. However, since every bundle contains an equal number of balls with gold and without gold, regardless of the *e* value, randomly extracting *e* bundles will yield an equal number of balls with gold and without gold. Therefore, if one has the notepad, by looking at it and selecting all colors containing gold, then summing all selected colors, the normal array yields a value equal to 1/2 of the total number of balls drawn, while the contaminated array yields a value different from 1/2. The fact that only the person with the notepad can distinguish between the normal and contaminated arrays means both arrays can be used as ciphertext. If the normal array is used as ciphertext '1' and the contaminated array as ciphertext '0', then the basket composed of ball bundles corresponds to the public key, and the notepad corresponds to the private key. Figure 1) shows a conceptual diagram of the proposed cryptographic algorithm.

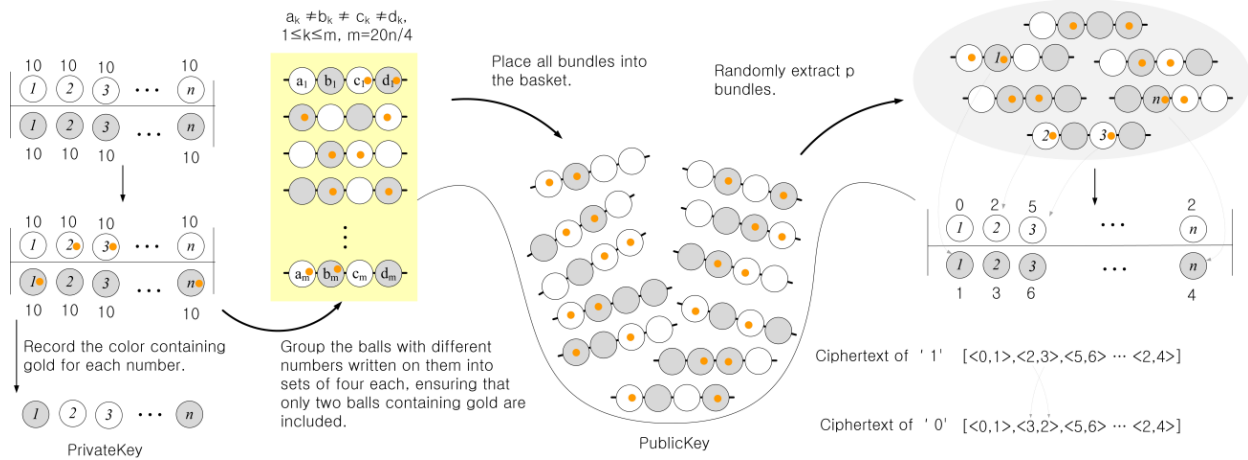

Fig. 1 Concept of the new public key based cryptography

When the bundles are unwrapped, the property that the number of balls containing gold equals the number without gold propagates to each individual ball, expressed as a spectrum representing the count of each ball. The number of distinct spectra is the combinatorial number $_{2mn/2k}C_e$. Since we can set *e* to any value smaller than *n*, choosing e=n/a (a>1, integer) exceeds the polynomial number for the input variable count *n*. Therefore, it is impossible to find the entire set of normal arrays within polynomial time, making it impossible to distinguish contaminated arrays from normal ones within polynomial time. We now describe how to concretize the above concept using the satisfiability problem (SAT).

**Materialization**

The **variables** constituting a SAT use Boolean variables that take only the values TRUE (1) or FALSE (0). **Literals** are represented as Boolean variables such as 'a' or the negation of a Boolean variable such as '¬a'. An expression where literals are connected by one or more disjunctions (∨, logical OR), such as (a ∨ ¬b ∨ c), is called a **clause**. An expression where clauses are connected by one or more conjunctions (∧, logical AND), such as (a ∨ ¬b ∨ c) ∧ (a ∨ c ∨ d), is called a **conjunctive normal form (CNF).** A CNF where all clauses consist of k literals is denoted as **k-CNF**. A clause is said to be **satisfied** if it contains at least one literal with a TRUE value. The set of literals with TRUE values that satisfies all clauses in a CNF is called a **solution** to the CNF, and the problem of determining whether a solution exists is called the **satisfiability problem**. From the perspective of the solution set, a clause consisting of k literals that are TRUE is defined as a **k-TRUE-clause**.

Replacing balls containing gold with literals with values of TRUE, and balls not containing gold with literals with values of FALSE, the set of balls is expressed as a clause. Since the bundles are restricted to contain exactly *k* balls containing gold when grouped into 2k bundles, all bundles inserted into the basket become k-TRUE-clauses. The basket thus corresponds to an SAT instance composed of a 2k-CNF with strengthened conditions for having a solution. To formalize the act of counting the number of balls containing gold into a mathematical expression, the following terms are defined.

Solution, *A*, contains *n* literals selected from the *2n* literals formed by the variables $x_k$(1≤k≤n). If $x_k$ is included in *A*, assign 1 to $x_k$ and 0 to ¬ $x_k$; if not included, assign 0 to $x_k$ and 1 to ¬ $x_k$. The resulting array of pairs ($x_k$, ¬ $x_k$) is defined as the **solution vector**.

$[(x_k, ¬x_k)]^n = [x_1, ¬x_1, x_2, ¬x_2, …, x_n, ¬x_n]$, $x_k$ and $¬x_k \in \{0,1\}$ …*(1)*

Generate the value of $a_k$ by counting the number of $x_k$ (1≤k≤n) in the multiset composed of literals included in CNF, generate the value of $b_k$ by counting the number of complements of $x_k$, and then create an array of $a_k$ and $b_k$ pairs as follows:

$[(a_k, b_k)]^n = [a_1, b_1, a_2, b_2, \ldots, a_n, b_n]$ …(2)

The generated array is defined as the **literal spectrum**. The formula for calculating the inner product between the solution vector *A* and literal spectrum *L*, where *L* is treated as a vector, is as follows.

$$L \cdot A = \sum_{k=1}^{n} \left(a_k x_k + b_k (\neg x_k)\right) \Big| L = [(a_k, b_k)]^n, A = [(x_k, \neg x_k)]^n, x_k \ \& \ \neg x_k \in \{0, 1\} \quad \cdots (3)$$

The value of Equation (3) represents the number of times the literal spectrum uses the literals contained in solution *A*. The number of literals contained in the solution also signifies the strength of satisfiability for CNF. Therefore, the above value is defined as a **satisfiability measure.**

The array recording the number of extracted balls becomes the literal spectrum, and the act of counting the number of balls containing gold by looking at the notepad becomes the act of calculating the satisfiability measure. Since only *k* literals with TRUE values are included in all clauses, extracting *e* clauses from 2k-CNF yields a satisfiability measure value of *ke* for the literal spectrum.

The procedure for generating a public-private key pair is as follows.

*Public key and private key generation procedure)*

1. Set the number of input variables to *n* and the multiplicity of the multiset to *m*.

2. Generate a solution vector of size *n* randomly and use it as the private key.

3. Divide the 2mn literals into groups based on the solution vector, where one group holds TRUE values and the other holds FALSE values.

4. Randomly extract *k* literals from each group, ensuring that the *2k* literals are distinct and that no literal is extracted together with its complement.

5. Repeat step 4 on the remaining TRUE and FALSE groups to generate mn/k clauses.

6. The generated 2k-CNF composed of mn/k k-TRUE-clauses is used as the public key.

After generating the public key and private key, the procedure for creating ciphertext and plaintext using the new cryptographic concept described above is summarized as follows.

*Method 1)*

Method 1) requires more computational effort to generate the ciphertext for plaintext '0' than for plaintext '1'. To prevent vulnerability to side-channel attacks due to this computational difference, the algorithm is modified as follows.

*Method 2)*

1. The ciphertext writer generates a literal spectrum by randomly extracting e clauses from the 2k-CNF composed of k-TRUE-clauses used as the public key.

2. When encrypting plaintext '1', randomly select one literal from the literal spectrum where the number of extracted counts for the literal and its complement are equal, then transmit it after swapping the extracted counts. When encrypting '0', randomly select one literal from the literal spectrum where the number of extracted counts for the literal and its complement are different, then transmit it after swapping the extracted counts.

3. During decryption, if the satisfiability measure value is *ke*, it is restored as 1; otherwise, it is restored as 0.

Figure 3) shows the block diagram of the proposed cryptographic algorithm.

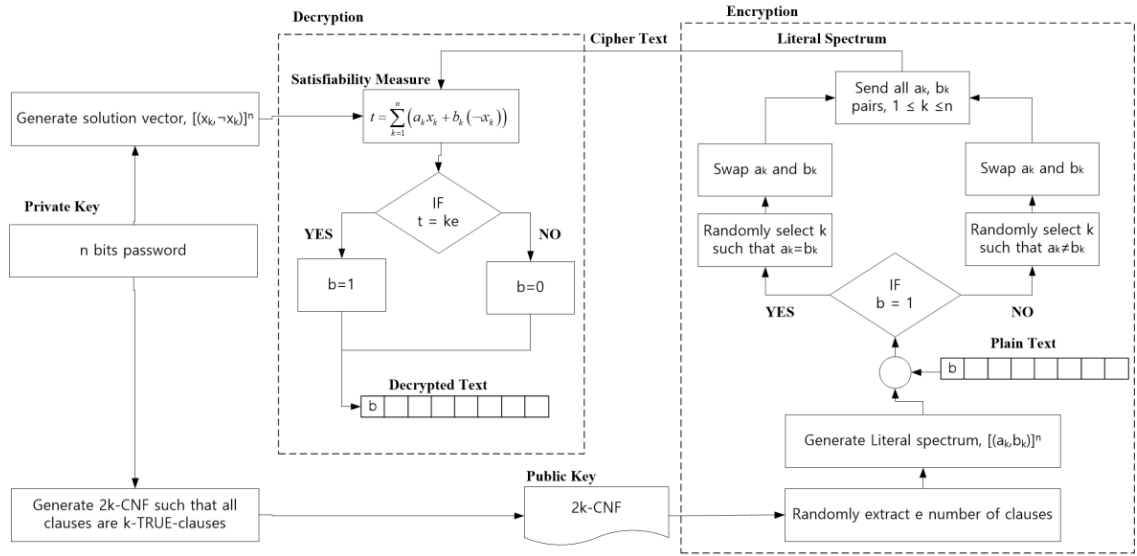

Fig. 3 Method 2: SAT based PQC for 1 bit plaintext considering side-channel attack.

What happens if we attempt restoration with an arbitrary password? Substituting an arbitrary password into the CNF generates clauses with a number of literals holding TRUE values that differs from the intended *k*. This paper refers to such clauses as contaminated clauses. Because sporadic contaminated clauses exist, there is no guarantee that the same literals with TRUE values and literals with FALSE values will always be extracted each time clauses are extracted. Consequently, it becomes impossible to distinguish between 1 and 0 using the satisfiability measure value $ke$.

The proposed method is very fast at generating ciphertext because it only requires randomly sampling and counting the number of literals. However, it has the disadvantage of increasing ciphertext length since the counts of all sampled literals must be recorded in the ciphertext. To mitigate this drawback, encryption is performed in units of b bits rather than 1 bit, as follows.

*Method 3)*

1. To generate ciphertext for b bits of plaintext, not 1 bit, randomly select $2^b$ variables from those that make up the public key.

2. Extract clauses such that the difference in the count values of the two literals formed by each of the selected $2^b$ variables ranges from 0 to $2^b$ - 1. Here, the value $2^b$ must be less than or equal to the multiplicity of the multiset.

3. The ciphertext writer divides the plaintext into b-bit segments. For each b-bit segment with value $p$ ($0 \leq p < 2^b$), swap the count values of the two literals created by the variable whose count value differs by $p$.

4. The ciphertext writer cannot determine whether swapping counts increases or decreases the satisfiability measure. Therefore, during decryption, if the satisfiability measure is ke+p or ke–p, $p$ is replaced with *b* bits to restore the plaintext.

This method slows encryption speed but reduces ciphertext length to 1/b by converting 1-bit ciphertext into b-bit units. It is used when short ciphertext length is critical; when fast encryption speed is critical, the previously described method is used.

Figure 4 shows the block diagram of the proposed cryptographic algorithm.

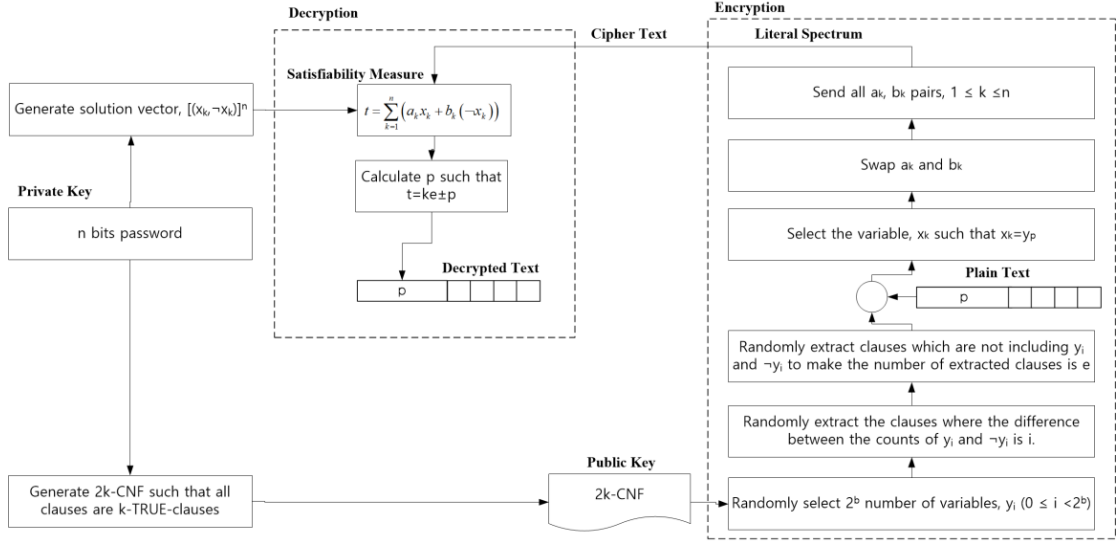

Fig. 4 Method 3: SAT based PQC for b bits plaintext

**Expansion to (2k+1)-CNF and cryptography using the divided public key**

Consider the case where the public key is divided into two groups. If we separate and publish the k-TRUE-clauses and the (k+1)-TRUE-clauses, the literals in the (k+1)-TRUE-clauses have a probability of being TRUE that is 1/2k higher than the literals in the k-TRUE-clauses. Therefore, by checking which group contains more instances of a literal and its complement, we can determine which group has a higher probability of being TRUE. However, when attempting to recover a 256-bit secret key based on probability without knowing the private key, if not all bits are correct, there is no way to verify whether the correctly recovered bits are indeed correct. Thus, high probability offers little help in block-level recovery. When splitting the public key into two groups, even when using (2k+1)-CNF, making half of all clauses k-TRUE-clauses and the other half (k+1)-TRUE-clauses maintains formal symmetry. This is because a single literal and its complement occur in equal numbers throughout the entire CNF.

To generate a (2k+1)-CNF composed of k-TRUE-clauses and (k+1)-TRUE-clauses, perform step 4 of the previously described public key and private key generation procedure: extract k literals from the group of literals with TRUE values and k+1 literals from the group of literals with FALSE values to create a k-TRUE-clause. Then, modify the process to create a (k+1)-TRUE-clause by extracting k+1 literals from the group with TRUE values and k literals from the group with FALSE values.

The procedure for generating ciphertext and decrypted text by splitting the public key into two groups is summarized below.

*Method 4)*

1. Divide the public key composed of (2k+1)-CNF into the k-TRUE-clause group and the (k+1)-TRUE-clause group.

2. Set the number of bits *b* in the plaintext to be encrypted and the number of clauses *e* to be extracted such that $2^b=e$.

3. To encrypt p (0≤p<e) values, extract *p* clauses from the k-TRUE-clause group and *e-p* clauses from the (k+1)-TRUE-clause group, and then generate the literal spectrum.

4. The satisfiability measure value t becomes kp + (k+1)(e-p). Since the values of *k* and *e* are known, the *p* value is recovered from the relationship p=(k+1)e-t.

By setting the *e* value to $2^{10}$ to $2^{12}$ based on the number of input variables and the multiplicity value, plaintext of 10 to 12 bits can be encrypted at once, reducing the ciphertext length by 2.5 to 3 times. Additionally, it offers the advantage of significantly shorter ciphertext generation time compared to Method 2).

Figure 5 shows the block diagram of the proposed cryptographic algorithm.

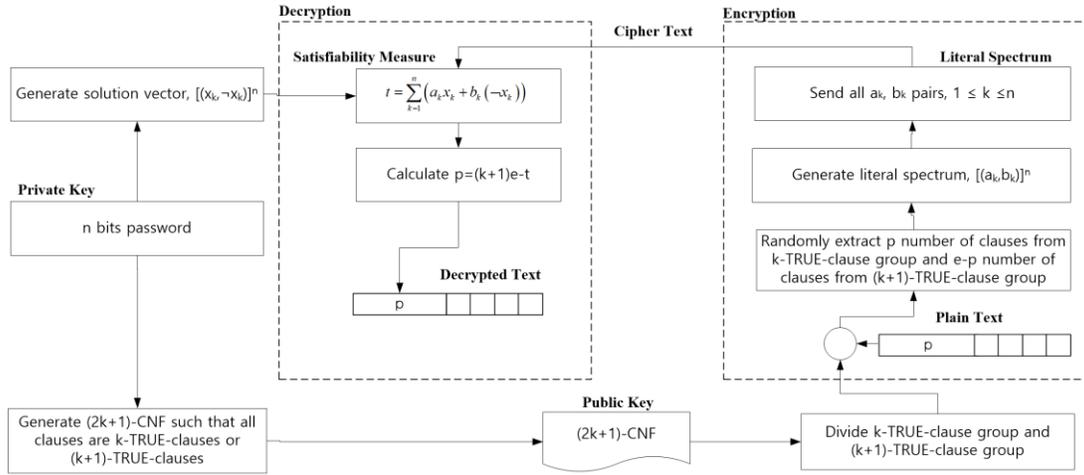

Fig. 5 Method 4: SAT based PQC using the divided (2k+1)-CNF

**System variables**

The system variables of the proposed cryptosystem are $k$, $b$, $m$, $n$, $e$ and $q$, each with the following meanings: k: Number of literals included in the clauses, b: Number of plaintext bits used during ciphertext generation, m: Number of times a single literal is used in the public key, representing the multiplicity value of the multiset, n: Number of input variables, e: Number of clauses randomly selected, q: Number of bits used to record the literal count value.

The table below shows an example of system variable settings for comparison with CRISTALS- Kyber.

| | c | ct (b bits) | sk | pk | ct (256 bits) |
|---|---|---|---|---|---|
| (k, b, m, n, e, q) | 2mn/k | 2qn | n | $2mn(\log_2 n+1)$ | ct * 32(byte)/b |
| (4, 4, 20, 512, 0.3*c, 4) | 5K | 4K | 64 (byte) | 25K (byte) | 32K (byte) |
| (4, 4, 20, 1024, 0.3*c, 4) | 10K | 8K | 128 (byte) | 55K (byte) | 64K (byte) |
| (5, 10, 20, 512, $2^b$, 4) | 4K | 4K | 64 (byte) | 25K (byte) | 12.8K (byte) |
| (5, 10, 20, 1024, $2^b$, 4) | 8K | 8K | 128 (byte) | 55K (byte) | 25.6K (byte) |
| CRISTALS-Kyber 1024 | | | 3168 (byte) | 1568 (byte) | 1568 (byte) |

Table 1 SAT based PQC using the divided CNF, c: number of clauses in public key, ct: cipher text size, sk: secrete key(private key) size, pk: public key size

Compared to CRISTALS-Kyber 1024, the length of the public key and ciphertext is approximately 10 to 40 times larger, but the length of the private key is about 25 to 50 times smaller. Furthermore, while CRISTALS-Kyber has a non-zero probability of decryption failure, the proposed algorithm achieves a decryption failure probability of zero by not using random values.

## DISCUSSION

Many public-key algorithms impose constraints to create one-way properties and then add a trapdoor, which has provided opportunities for various attacks. Lattice-based and code-based algorithms create one-way properties by inserting noise, and then add a trapdoor. For lattice-based algorithms, this is limited to Gaussian or small noise, while for code-based algorithms, the weight of noise-containing codewords is restricted to be below a specific value. Algorithms using the knapsack problem also exploit its one-way property, but the constraint of a superincreasing sequence imposed to add the trapdoor [19] provided the basis for subsequent polynomial-time attack algorithms [20]. The algorithm proposed in this paper utilizes the one-way property between the subsets of a multiset and the number of elements. It adds a trapdoor by ensuring the number of literals with TRUE values in the subset is fixed. However, since finding the private key using this property requires search algorithms rather than mathematical analysis, it is immune to various algebraic attacks. Furthermore, the conditions for a solution to 2k-SAT are strengthened such that exactly K literals must be TRUE, whereas the original condition required at least one TRUE literal in every clause. This significantly increases the time SAT solvers require to find a solution compared to standard SAT problems.

The proposed cryptographic algorithm is quantum resistant because it does not rely on the difficulty of integer factorization or the discrete logarithm problem. the proposed algorithm is expected to serve as a new shield against quantum computing.

# REFERENCES AND NOTES

[1] P. W. Shor, "Algorithms for quantum computation: discrete logarithms and factoring," in Proc. 35th Annual Symposium on Foundations of Computer Science, 1994, pp.124-134.

[2] P. W. Shor, Polynomial-time algorithms for prime factorization and discrete logarithms on a quantum computer. SIAM Rev. 41, 303–332 (1999).

[3] A. D. King, J. Carrasquilla, J. Raymond, I. Ozfidan, E. Andriyash, et al., "Observation of topological phenomena in a programmable lattice of 1,800 qubits," Nature 560 (2018).

[4] D-Wave Systems reports on quantum processor in Nature Magazine, Nature, 2011.05.12

[5] F. Arute, K. Arya, R. Babbush, et al. "Quantum supremacy using a programmable superconducting processor," Nature, 574, 2019, pp.505-510.

[6] R. Avanzi, et al., "CRYSTALS-Kyber: Algorithm Specifications And Supporting Documentation," NIST PQC Round 3 submission, Oct. 1, 2020.

[7] C. Chen, et al., "NTRU: Algorithm Specifications And Supporting Documentation," NIST PQC Round 3 submission, Sep. 30, 2020.

[8] A. Basso, et al., "SABER: Mod-LWR based KEM(Round 3 Submission)," NIST PQC Round 3 submission, Oct. 21, 2020

[9] M.R. Albrecht, et al., "Classic McEliece: conservative code-based cryptography," NIST PQC Round 3 submission, Nov. 19, 2020.

[10]. Melchor, C.A., et al, Nist post-quantum cryptography standardization round 4 submission: Hamming Quasi-Cyclic (HQC)

[11] L. G. Valiant *et al.* NP is as easy as detecting unique solutions Theor. Comput. Sci. (1986)

[12] S. Cook, The complexity of theorem proving procedures. in *Proceedings of the 3rd Annual ACM Symposium on Theory of Computing* (Association for Computing Machinery, New York, 1971), p. 151.

[13] Sebastian E. Schmittner, A SAT-based Public Key Cryptography Scheme, arXiv:1507.08094 [cs.CR]

[14] Republic of Korea Patent No. 10-2837502: Method for Implementing a Quantum-Resistant Cryptographic Algorithm Based on a Satisfiability Problem with Enhanced Conditions for Having a Solution

[15] PCT/KR2025/095085: A Method for Implementing a Quantum-Resistant Cryptographic Algorithm Based on a Satisfiability Problem with Enhanced Conditions for Having a Solution

[16] Republic of Korea Patent No. 10-2657596: Method for Generating Hard SAT and Method for Implementing a Quantum-Resistant Cryptographic Algorithm Based on SAT

[17] Republic of Korea Patent No. 10-2762363: Method for Implementing a Quantum-Resistant Cryptographic Algorithm Based on SAT

[18] W. DIFFIE, M. E. HellMAN, New Directions in Cryptography IEEE TRANSACTIONS ON INFORMATION THEORY, VOL. IT-22, NO. 6, NOVEMBER 1976

[19] Merkle, Ralph; Hellman, Martin (1978). "Hiding information and signatures in trapdoor knapsacks". IEEE Transactions on Information Theory. **24** (5): 525–530. *doi*:*10.1109/TIT.1978.1055927*.

[20] Shamir, Adi (1984). "A polynomial-time algorithm for breaking the basic Merkle - Hellman cryptosystem". IEEE Transactions on Information Theory. **30** (5): 699–704. doi:10.1109/SFCS.1982.5.